# Roughness enhanced K x-ray emission from femtosecond laser produced copper plasmas

**Paramel P. Rajeev, Arvinder S. Sandhu and G. Ravindra Kumar**[*]
*Tata Institute of Fundamental Research, Homi Bhabha Road, Bombay 400 005, India*

## Abstract

K x-ray emission from copper plasmas produced by 100 fs, 806 nm laser pulses at intensities in the range of $10^{15}$-$10^{16}$ W cm$^{-2}$ is studied with respect to input laser polarization and target surface conditions. We demonstrate that even moderate surface roughness enhances the characteristic x-ray yield by as much as 70 %. We further show that surface roughness of the targets overrides the effect of polarization state in light coupling to the plasma.

PACS : 52.25.Nr, 52.40.Nk, 52.50.Jm, 42.65.Re

---

[*] E-mail: grk@tifr.res.in. Fax: +91-22-215 2110. Author for correspondence

Solid plasmas produced by intense, femtosecond lasers offer tremendous promise as `point' (micron sized) sources of ultra short x-rays[1-4], particularly for applications like lithography and time resolved diffraction. Methods to enhance x-ray yield are of obvious importance, and influence of various laser and target conditions is being investigated. Significant enhancement in yields is reported using pre-formed plasmas [3,4] though the x-ray pulse duration becomes longer[3]. Recent investigations report on the role of modulation of target surface in enhancing x-ray yield. Impressive enhancements of x-ray flux have been achieved in nanohole alumina targets (soft x-ray region) [5] and nickel 'velvet' targets (soft-hard x-ray region) [6]. Most of the previous work has examined enhancements in the soft x-ray region. In our recent studies on x-ray emission, we observed that inexpensive, unpolished copper targets showed significant enhancement in yields in both hard and very hard x-ray regimes[7] compared to polished ones. The K x-ray of copper produced by femtosecond laser excited copper plasmas, has in particular been used to demonstrate real time x-ray diffraction imaging[8,9] on ultrashort timescales and is considered a source with great potential for applications in condensed mater physics and biology. Any attempts to enhance such emission are therefore highly desirable. In this letter, we discuss the role of surface roughness of the targets in enhancing the characteristic emission from copper targets and make the crucial point that roughness overrides the effect of input polarization in the generation of K x-rays.

Our experiments are done using a Ti: Sapphire chirped pulse amplified laser emitting 100 fs pulses at 806 nm. The laser pulses have a contrast of $10^5$ with the baseline and under optimum extraction, the prepulse intensity level (13 ns before) is less than $10^{-4}$ of the main pulse. The prepulse or the pedestal do not cause significant plasma formation. The laser is focused at oblique incidence (45°) with a 30-cm focal length lens on copper targets housed in a vacuum chamber at $10^{-3}$ torr. The maximum pulse energy used in the experiments is 3 mJ, giving a peak intensity of about $1 \times 10^{16}$ W cm$^{-2}$ at a focal spot size of 30 μm. A thin half wave plate in the beam path selects the polarization state. The target is constantly rotated and translated in order to avoid multiple hits at the same spot by the laser pulses. X-ray emission from the plasma is measured along the plasma plume direction using a cooled silicon detector (Amptek, XR-100CR). The chamber has a 25μm thick mylar window, which sets a lower energy cutoff of about 1 keV for the observed emission. Spectra were typically collected over 30000-40000 laser shots.

Targets studied had an 'average roughness' (AR) of 120 nm, 18 nm and 5 nm (Fig.1 has AFM images of two of them). The first was a commercially available unpolished sample, while the other two were polished to different degrees. Fig. 2 (a) shows the K x-ray spectra for s-polarization of the laser. The $K_\alpha$ and $K_\beta$ lines are clearly resolved for all the targets. The roughest target gives a yield 70% higher than the smoothest (polished) one. The roughest target has the $K_\alpha$ emission appearing at 8.05 keV and $K_\beta$ at 8.89 keV, while the smoothest target has them appearing at 8.53 keV and 9.39 keV and the target with intermediate roughness has peaks at 8.4 keV and 9.3 keV. In other words, the smoothest target has the largest blue shift for the peaks. The spectral widths are 230 eV for $K_\alpha$ while they are 340 eV for $K_\beta$. These are clearly larger than the detector resolution of 190 eV, indicating the presence of a wide range of charge states. The $K_\alpha / K_\beta$ peak intensity ratio is 6:1 which is close to the expected value[10].

Figure 2(b) presents x-ray spectra for s and p polarizations for the target with average roughness 120 nm. The spectra for both polarizations look nearly identical implying that polarization does not have any impact on the characteristic emission from a target with a reasonable amount of roughness. These observations provoke questions on the role of roughness vis-à-vis the polarization of the laser, in the generation of the K emission. For p-polarization, obliquely incident laser beam generates hot electrons via resonance absorption (RA) [11] and vacuum heating[12] and these hot electrons generate characteristic K emission from the cold, bulk target behind critical layer[13-15]. Thus the presence of hot electrons is a key factor in the emission process, which depends on the transport of hot electrons and their radiation through the plasma layer to the bulk. The surface topography may thus have a significant role to play in the transport process. In fact, earlier studies have shown inhibition of electron transport into porous targets[16].

For hot electrons produced by RA, the scaling law[17] gives a temperature $T_{hot} = 14\ T_c^{0.33} (I\lambda^2)^{0.33}$, where $T_c$ is the background electron temperature in keV, I is the intensity of the laser in units of $10^{16}$ W cm$^{-2}$ and ë is the wavelength in microns . For a $T_c$ of 0.1 keV, we get a $T_{hot}$ of 6.6 keV under our experimental conditions, which is close to one of the components we observed (reported elsewhere)[7]. In addition, we had also observed a component with a temperature of about 40 keV [7]. On the contrary, in the case of s-polarized light field, there is no known mechanism other than inverse bremsstrahlung operative under our experimental conditions and even this is expected to be less effective[11] above $10^{15}$ W cm$^{-2}$. Surprisingly, the integrated x-ray emission from unpolished targets is almost equal for both polarizations despite the fact that the coupling should be much weaker using s-polarized light. These observations demand an examination of the role of roughness in the generation of hot electrons. The understanding is that structuring of the surface leads to localized volume heating of micro regions of the target resulting in denser plasmas and higher temperatures[5,6].

We obtain a higher K x-ray yield, but almost no blue shift for the roughest target. Enhanced yield can be directly attributed to (1) higher number of hot electrons, (2) their confinement by the roughness of the target surface and hence increase in their interaction with unionized target atoms (equivalent to penetration through more bulk) and (3) more surface area available for the x-ray emision due to the roughness. The lack of any blue shift in the roughest target indicates that the emission arises mainly from atoms with a single vacancy, in contrast to smooth targets. The charge states in the case of the smoothest targets can be estimated to be in the range of 25+ to 28+ corresponding to a blue shift of 350 - 480 eV. This may be explained as follows. In a smooth target, the hot electrons could go deep into the bulk to cause K x-ray emission and there may be considerable reabsorption or scattering of the resulting K x-rays. Emission observed would be mainly from the thin expanding plasma layer on the surface. This layer consists mostly of ionized atoms of varying charge states, giving rise to the blue shift. On the contrary, in rough targets not all neutral atoms may be ionized equally (the surface irregularities may enhance the field at certain points and reduce it at other positions), particularly those in valleys may not be highly ionized. Thus there could be a dominance of singly charged ions with a single K shell vacancy. In addition, even for ions with higher charge, the confinement effects of roughness may increase local densities. This in turn enhances electron-ion recombination[18] leaving a predominant population of singly ionized atoms with K shell vacancies, which could give unshifted characteristic emission.

The polarization independent K x-ray emission from unpolished targets could mainly be attributed to two factors – (i) the lack of a definite geometry on the surface because of which polarization could be considered either s or p depending on the local topographical feature at the point of incidence, and (ii) modification of polarization by scattering from the rough features.

These factors could imply that the influence of roughness could override the role of polarization in light coupling to the plasma. We note that Ahn et al.[19] have seen similar lack of dependence of soft x-ray yield on light polarization state and cited rippling of the critical surface[20] as a possible cause.

Interestingly there is a wide variation in the copper K x-ray yields reported in different experiments as well as predicted by simulations, for experimental conditions similar to ours. Experimental reports have quoted $10^{-7}$ % ($10^2$ photons/sr/pulse) at an incident intensity of $2 \times 10^{15}$ W cm$^{-2}$ of KrF laser pulses[21] and $5 \times 10^{-3}$ ($10^{10}$ photons/sr/pulse) using a Ti-sapphire (780 nm), $3 \times 10^{16}$ W cm$^{-2}$ laser[22]. Guo et al[23] have reported yields similar to the latter study. Among predictions, Reich et al [24] obtained an optimum $10^{10}$ photons/sr/pulse for copper at $2 \times 10^{16}$ W cm$^{-2}$, while Rousse et al [15] report only $10^6$ photons/sr/pulse. In our study we estimated a maximum yield of $10^4$ photons/sr/pulse (roughest target). It is thus important to note that the yield may be critically dependent on actual experimental conditions like focusing and type of target used.

In conclusion, we have demonstrated that the unpolished, readily available and inexpensive surfaces could be used to produce larger fluxes of ultrashort K x-ray pulses than polished ones. A key observation is the lack of influence of the polarization state on the x-ray (hot electron) yields in the case of unpolished targets in stark contrast to the observations for polished targets. Further experiments are under way to study the enhanced K x-ray generation in targets with tailored roughness.

We thank S.P. Pai and J. John for the AFM images and Sudip Sengupta, P.K. Kaw, V.R. Marathe and M. Krishnamurthy for discussions. The high energy, femtosecond laser facility has received substantial funding from the Department of Science and Technology, Government of India, New Delhi.

___________________________________________________________________________

**Figure Captions:**

1. Atomic Force Microscope (AFM) images of two targets studied (a)- "average roughness (AR)" = 5 nm (polished); (b) AR=120 nm (unpolished).

2. (a) K x-ray emission from targets with different average roughness (AR) for s-polarized laser light. (b) X-ray emission for s and p-polarized laser light from an unpolished target

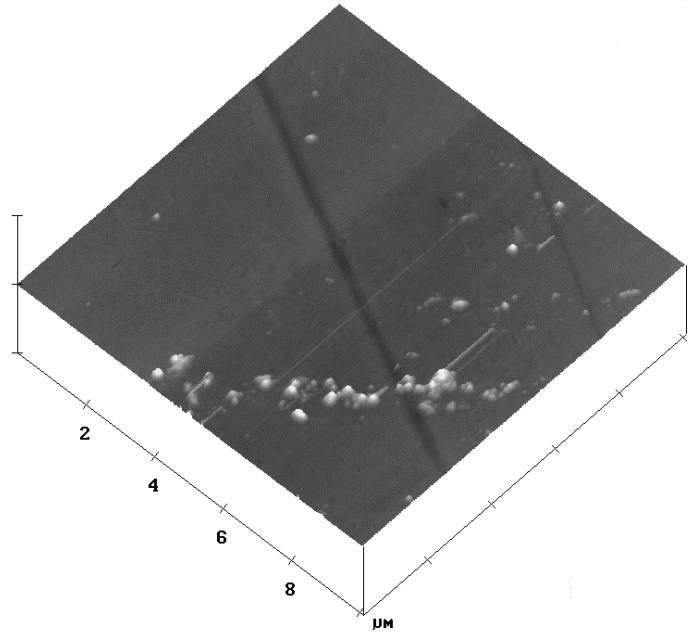

(a)

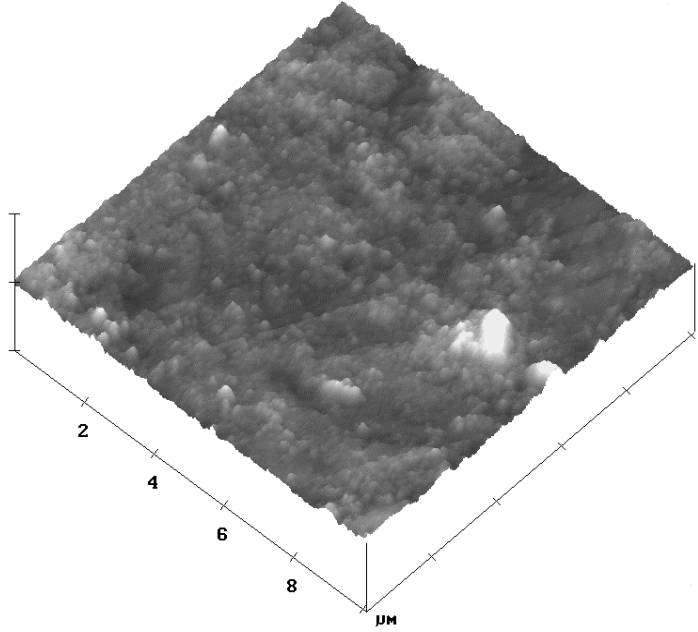

(b)

Fig1: P.P. Rajeev et. al

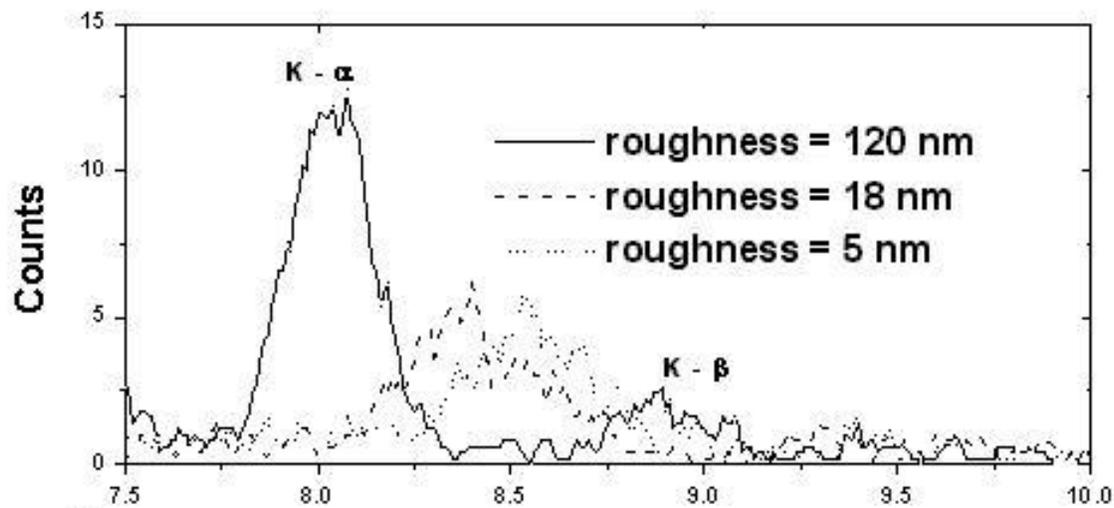
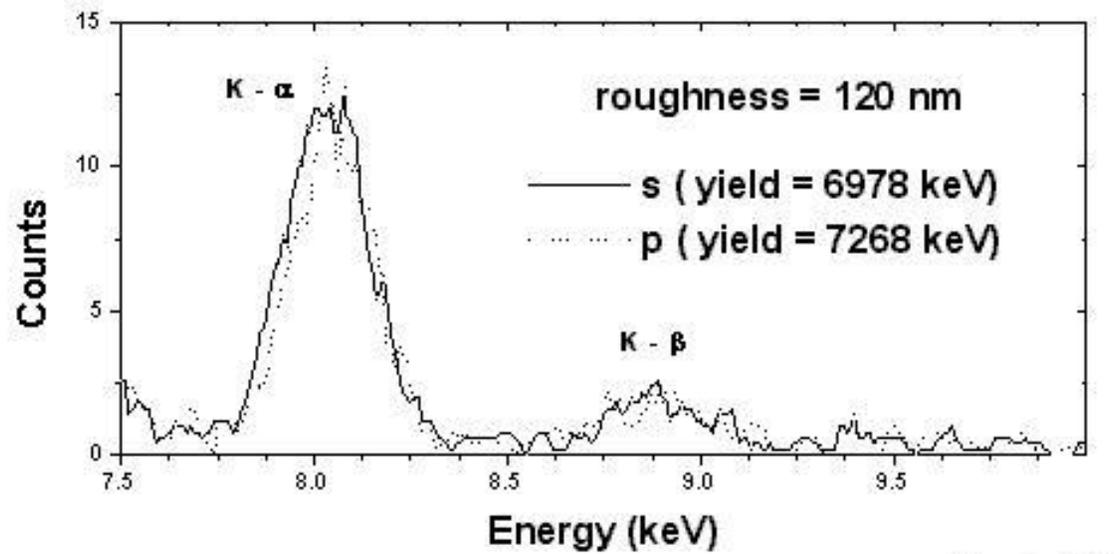

Fig 2: P.P. Rajeev et. al.